\documentclass[a4paper,11pt]{article}

\usepackage{mathrsfs}

\title{\bf $\Theta$-parity in supersymmetry}
\author{\bf Pritibhajan Byakti and Palash B Pal\\
Saha Institute of Nuclear Physics\\
1/AF Bidhan-Nagar, Kolkata 700064, India.}
\date{}

\def\ms#1{\mathscr #1}
\def\Eqn#1{Eq.\,(\ref{#1})}

\begin{document}

\maketitle

\begin{abstract}
  We consider invariance of the action of $N=1$ supersymmetric
  theories under the change of sign of the fermionic co-ordinate in
  superspace.  We show that the $R$-parity can be realized as a
  special implementation of this symmetry.  Other implementations in
  the supersymmetric extension of the standard model can be related to
  lepton number and baryon number parities.

\end{abstract}

\bigskip

The most general Lagrangian in any realistic supersymmetric theory
contains a huge number of interaction terms.  From time to time,
various discrete symmetries have been proposed to cut down on the
number of terms and to meet phenomenological constraints.  The
matter-parity \cite{Mparity} was introduced, assuming it to be
negative for superfields containing quark and lepton fields and
positive for others, in order to prohibit renormalizable couplings
which violate baryon and lepton numbers.  The $R$-parity, equal to
$(-1)^{3B+L+2S}$ on spacetime fields where $B$, $L$ and $S$ are the
baryon number, lepton number and spin of any particle, was introduced
\cite{Farrar:1978xj} to ensure that the superpartners of ordinary
particles can be produced only in pairs.  It was subsequently pointed
out \cite{Hall:1983id, Pal:1998un} that all physical consequences of
these two symmetries are identical.

Here we show that the $R$-parity, and some similar symmetries whose
physical content will be discussed later, can arise from a much
general consideration.  The idea also does better justice to the word
``parity'', which was originally used to talk about space inversion.
Supersymmetric theories are best described in superspace, where the
usual spacetime co-ordinates are augmented by some spinorial
co-ordinates.  In simple, i.e., $N=1$ supersymmetry, there is only one
spinor co-ordinate of this sort.  We will call it $\theta$, and,  in
analogy with space inversion, consider inversion of its components:
\begin{eqnarray}
\theta \to -\,\theta \,.
\end{eqnarray}
This operation will be called $\Theta$-parity.

If all superfields are invariant under this operation, this is clearly
and trivially a symmetry of the action of any supersymmetric theory,
renormalizable or not, because any term in the action will contain
either integration over all four components of $\theta$, or on the two
independent components of a chiral projection of $\theta$.  More
explicitly, the action of an $N=1$ supersymmetric Yang-Mills theory is
of the form
\begin{eqnarray}
\ms A = \int d^4x \bigg( d^4\theta \; \ms K + \Big( d^2\theta_L \; \ms
W + \mbox{h.c.} \Big) + \mbox{(pure gauge terms)} \bigg)  \,.
\end{eqnarray}
Here $\ms K$ is the gauge invariant K\"ahler potential and $\ms W$ is
the superpotential.  In order to keep the gauge invariant
K\"ahler term invariant under $\Theta$-parity, we must require
\begin{eqnarray}
\Theta V(x,\theta) \Theta^{-1} = V(x,-\theta) \,,
\label{V}
\end{eqnarray}
which obviously keeps the pure gauge terms invariant as well.
However, chiral scalar superfields can have a non-trivial
transformation property.  The general rule for their transformation
would be
\begin{eqnarray}
\Theta \Phi(x,\theta) \Theta^{-1} = \eta_\Theta \Phi(x,-\theta) \,,
\end{eqnarray}
where $\eta_\Theta$ can be called the intrinsic $\Theta$-parity of the
superfield.  Since $\Theta$-parity, applied twice, produces the
identity operation, we must have $\eta_\Theta=\pm1$ for any
superfield.  \Eqn{V} can also be stated by saying that the intrinsic
$\Theta$-parity of the real scalar superfields must be $+1$.

We now examine what are the possibilities of intrinsic
$\Theta$-parities of different chiral superfields of the MSSM.  We
start with the following terms in the superpotential:
\begin{eqnarray}
\ms W_1=y_u U^c Q H_u + y_d D^c Q H_d + y_e E^c L H_d+\mu H_u H_d \,.
\label{W1}
\end{eqnarray}
The notation is standard: $Q$ and $L$ denote quark and lepton doublets
respectively; $U^c$, $D^c$ and $E^c$ are the complex conjugates of the
right handed singlets; and $H_u$, $H_d$ are the Higgs superfields.
All of these are left-chiral superfields.  Gauge and generation
indices have been omitted.  There are other terms which are allowed by
the gauge symmetry and supersymmetry, viz.,
\begin{eqnarray}
\ms W_2 &=& \lambda LLE^c + \lambda' LQD^c + \mu' LH_u \,, 
\label{W2} \\ 
\ms W_3 &=& \lambda'' U^c D^c D^c \,.
\label{W3}
\end{eqnarray}
However, the terms in \Eqn{W1} are crucial in the sense that they are
responsible for the masses of the quarks, the charged leptons and the
Higgs bosons.  So we want to start with them while discussing the
consequences of $\Theta$-parity.

If we consider only one generation of fermions, there are seven
superfields that appear in \Eqn{W1}.  There are four terms there,
which imposes four conditions on the intrinsic $\Theta$-parities of
the superfields.  So we can take three intrinsic $\Theta$-parities
independently.  There will be eight such combinations then.  Let us
list the ones for which $\eta_\Theta(H_u) = +1$.  Note that the
intrinsic $\Theta$-parities of $H_u$ and $H_d$ superfields must be the
same because of the $\mu$-term.  Hence, the list is as follows:
\begin{eqnarray}
  \begin{array}{c|cccccc}
    \hline
& \multicolumn{6}{c}{\mbox{Superfield}} \\ \cline{2-7}
\eta_\Theta & H_u, H_d & Q & L & U^c & D^c & E^c \\ 
\hline
\mbox{Choice 1} & +1 & +1 & +1 & +1 & +1 & +1 \\
\mbox{Choice 2} & +1 & +1 & -1 & +1 & +1 & -1 \\
\mbox{Choice 3} & +1 & -1 & +1 & -1 & -1 & +1 \\
\mbox{Choice 4} & +1 & -1 & -1 & -1 & -1 & -1 \\
\hline
  \end{array}
\label{choices}
\end{eqnarray}
The list will remain the same even if we consider multiple generations
of fermions, provided we insist that all fermion generations transform
identically under the $\Theta$-parity.

There are four more assignments of $\Theta$-parities of the fields
which keep all terms of \Eqn{W1}.  These can be obtained by reversing
the sign of the intrinsic $\Theta$-parities of all doublets of the
weak SU(2), not disturbing those of the SU(2) singlets.  However, it
can be easily seen that the resulting intrinsic $\Theta$-parities are
just $(-1)^{6Y}$ times the entries appearing in the list above, where
the weak hypercharge $Y$ is normalized in such a way that it is equal
to the electric charge for SU(2) singlets.  Thus, these are not new
possibilities: they arise as combination of gauge symmetry and the
possibilities listed in \Eqn{choices}.

Let us now examine the physical consequences of the different choices
catalogued in \Eqn{choices}.  Choice 1 does not produce any constraint
on possible allowed terms in the Lagrangian.  Thus, anything that is
allowed by gauge symmetry and supersymmetry is allowed by this
choice.  The others imply restrictions, as follows:
\begin{eqnarray}
\mbox{Choice 2} &:& \ms W_2 = 0 \,, \\
\mbox{Choice 3} &:& \ms W_3 = 0 \,, \\
\mbox{Choice 4} &:& \ms W_2 = \ms W_3 = 0 \,.
\end{eqnarray}
Note that the terms in $\ms W_2$ are lepton-number violating, so that
Choice 2 is equivalent to a lepton number parity $(-1)^L$.  Similarly,
$\ms W_3$ contains baryon-number violating terms, so that Choice 3 is
equivalent to a baryon number parity $(-1)^{3B}$.  If we take Choice
4, on the face of it, looks exactly like matter parity.  As already
said, once this is imposed, no baryon-number or lepton-number
violating term is present in the superpotential.  This is equivalent
to imposing $R$-parity, or $(-1)^{3B+L}$, on the superfields.  Long
ago, Hall and Suzuki~\cite{Hall:1983id} presented the idea of
$R$-parity in this manner.  However, they were not interested in the
possibility of implementing $\Theta$-parity as a fundamental symmetry,
and did not consider the other choices that we have described,

Let us now examine the consequence of introducing right-chiral
neutrino fields in the model.  Terms of the form $LN^cH_u$ will be
allowed for any of the variants of $\Theta$-parity by a suitable
choice of $\eta_\Theta(N^c)$.  Such terms can produce Dirac mass terms
for neutrinos once the gauge symmetry is broken.  In addition,
Majorana mass terms of the $N^c$ fields will also be allowed.  So the
see-saw mechanism can also work.

We have so far described everything in terms of superfields.  The same
description can be given in terms of component spacetime fields.  For
chiral scalar superfields, the scalar field comes as the
$\theta$-independent term in the component field expansion, and
therefore should have the same intrinsic $\Theta$-parity as the
superfield.  The fermionic fields are accompanied with one power of
$\theta$ in the component field expansion, and therefore should have
the opposite intrinsic $\Theta$-parity.  Thus, in terms of component
fields, we can write the intrinsic $\Theta$-parity of the component
fields as $(-1)^{L+2S}$ for our Choice 1, as $(-1)^{3B+2S}$ for Choice
2, and as  $(-1)^{3B+L+2S}$ for Choice 3.  Any of these forms is also
applicable to the gauge fields and gauginos.

We should mention that phenomenological implications of
$\Theta$-parity is no different from that of a lepton-parity, a
$3B$-parity, or an $R$-parity.  What we show here is that these three
kinds of parities, i.e., $Z_2$ symmetries, can be motivated by some
discrete operation on the fermion co-ordinates in the superspace.  In
this sense, they all have the status of space inversion and time
reversal symmetries, and can have a natural implementation in
super-spacetime.

We thank Gautam Bhattacharyya for discussions.


\end{document}